\documentclass{Rinton-P9x6}

\begin{document}

\title{Searching for string theories  of the standard model}

\author{David Bailin} 
\address{Centre for Theoretical Physics, 
University of Sussex, Brighton BN1 9QJ, UK, D.Bailin@sussex.ac.uk, }

\author{George Kraniotis}
\address{Centre for Theoretical Physics, University of Sussex, 
Brighton BN1 9QJ,UK, G.Kraniotis@sussex.ac.uk}

\author{Alex Love}

\address{Department of Physics, Royal Holloway and Bedford New College, \\
University of London, Egham, Surrey TW20 0EX}


\maketitle

\abstracts{We briefly review attempts to construct string theories that yield
the standard model, concentrating on models with a geometric 
interpretation. Calabi-Yau compactifications are discussed in the context 
of both the weakly coupled heterotic string and the strongly
 coupled version, heterotic M-theory. Similarly, we consider orbifold
  compactifications of weakly coupled heterotic theory and of Type II
theories  in the presence of D-branes. The latter allows a ``bottom-up" 
construction that so far seems the most natural and direct
 route to the standard model. }

\section{Introduction and background}
The discovery by Green and Schwarz\cite{GS} that Type I string theory, with gauge group $SO(32)$, is free of gauge, 
gravitational and mixed anomalies 
sparked the first string revolution. The construction of the heterotic string theories\cite{GHMR},
 with associated gauge group $SO(32)$ or $E_8 \times E_8$, followed shortly after. 
 Since it was the first fully consistent quantum theory of gravity,
 and came equipped with a gauge group, it naturally 
suggested that string theory might have a r\^{o}le as a quantum theory of all of the interactions 
of nature. The immediate obstacle that the consistency of the theory requires a space-time with dimension 
$D=10$, whereas nature apparently has only $d=4$ dimensions, may be surmounted by 
requiring that the unobserved (six, assumed spatial) dimensions are compactified, in the manner 
first suggested by Kaluza and Klein over eighty years ago. It was next shown\cite{CHSW} that if we 
require that the $d=4$ dimensional Minkowski space-time has $\mathcal{N}=1$ supersymmetry, then there is a covariantly constant spinor, and
the compactified dimensions must form a space with $SU(3)$ holonomy group. This occurs if and only if it is a Ricci-flat K\"{a}hler 
manifold. Manifolds with this property are called Calabi-Yau manifolds, and in \S 2 we review briefly 
the attempts to find Calabi-Yau threefolds which yield a realistic gauge group and matter content.

The simplest compact flat space is just a torus $T^6$, but because it is Riemann- as well as Ricci-flat, the 
residual supersymmetry is $\mathcal{N}=4$. However, this can be reduced to $\mathcal{N}=1$ if we quotient the space by a 
``point group'' $P$ which acts crystallographically on the lattice of the torus. If $P$ has fixed points, the resulting space 
has curvature singularities at the fixed points, but is flat at all other points. Such spaces are called ``orbifolds'', and 
despite the singularities
it turns out that string propagation is well-defined on an orbifold\cite{DHVW}. Furthermore, all quantities in the effective
supergravity theory that emerges are much more readily calculable than in Calabi-Yau threefolds. It is therefore natural to 
to see whether orbifold compactifications can yield  a realistic gauge group and matter content, and the work on this 
topic is reviewed in \S 3.

Both of the foregoing attempts to extract realistic phenomenology from (the weakly coupled heterotic) 
string theory make explicit use of the geometry of the six unobserved spatial dimensions. However, the heterotic 
string is  a hybrid, whose right-moving degrees of freedom are those of the ($D=10$) Type II string, 
but whose left-movers are those of the original ($D=26$) bosonic string. The 
sixteen additional left-movers are regarded as ``internal'' degrees of freedom, not realised as  physical 
spatial dimensions, and are either compactified on a maximal torus or alternatively fermionized; six of the remaining nine spatial
 dimensions are then compactified in the ways described above.
 This suggests the possibilty that {\it none} of the six additional spatial dimensions is 
realised physically, and that they could all be regarded as internal degrees of freedom, like the sixteen superfluous 
left movers.  The need for $D=10$ space-time dimensions in superstring theory derives from the requirement that the 
total central charge of the world-sheet superconformal field theory vanishes, and this can be achieved by 
degrees of freedom that are not necessarily those that derive from geometric dimensions. This observation 
motivated the Gepner models\cite{Gepner}, in which the nine units of central charge,
 supplied to the superstring by the six compact (superspace) dimensions, are provided instead by various
  minimal $\mathcal{N}=2$ superconformal coset theories. Similarly, in the free-fermion models\cite{KLT,ABK}, all of 
the compact bosonic coordinates are fermionized, giving twenty real world-sheet fermionic right movers,
and 44 real world-sheet fermionic left movers. Different theories result from imposing different boundary conditions 
on these fermions,  and it is arguable that the most realistic string models found to date have been constructed in this way\cite{Faraggi}. 
However, the second string revolution, made great use of the geometric aspects of string theory, and so far as we are aware 
there have been no echoes of the non-geometric compactifications in the many new forms of string theory which ensued.
 We have therefore confined this brief review of the possibility of extracting realistic phenomenology from string theory to 
 the scenarios with a geometric interpretation.

The second revolution has led to the belief that the five known superstring theories are actually limiting cases 
of a deeper, more fundamental theory, M-theory, whose  precise nature has not yet been unravelled\cite{Witten}. The new insights have resolved several 
longstanding criticisms of (the theory formerly known as) string theory. For example, it has been known for many years that 
the largest space-time dimension of a supergravity theory is $D=11$, whereas all of the superstring theories utilize only $D=10$ dimensions. 
It seemed odd that the larger symmetry was not realized by any of these allegedly fundamental theories. Similarly, the power of string theory to 
tame the notorious divergences of quantum gravity derives from abandoning the requirement of (point-like) particle fundamental entities in 
favour of (one-dimensional) strings. This raises the question of whether (two-dimensional) membranes, or even higher ($p$-dimensional) 
$p$-branes might arise, especially since $p=2$ super-membranes\cite{HLP} were known to arise as solitons in $D=11$ supergravity\cite{BST}. In addition,
 the $D=10$ supergravity arising from the $E_8 \times E_8$ heterotic string has 5-brane solitons, whose mass per unit 5-volume becomes larger 
 as the string coupling gets weaker; this suggests a weakly coupled 5-brane when the string is strongly coupled. Following the revolution 
 it is now believed that in some region of moduli space ({\it i.e.} the vacuum expectation values that determine the strength of 
 the string coupling, and the size and shape of the compactified space), M-theory is well-approximated by $D=11$ supergravity,
  and that all of the five (perturbative) superstring theories arise as limiting cases in other regions. For example, if one coordinate ($x^{10}$) of
   the $D=11$ supergravity
 is compactified on a circle of radius $R$, which is wrapped by one of the (two) membrane dimensions, then when $R$ 
is small the pipe-like super-membrane world volume becomes that of a string, and the dimensionally-reduced 
supergravity action becomes that of the (non-chiral, $D=10$) Type IIA superstring\cite{DHI}. Likewise, the $E_8 \times E_8$ heterotic
 string may be obtained by compactification on a circular orbifold $S^1/Z_2$; the chiral matter arises from the need to cancel anomalies 
 at the orbifold fixed points at $x^{10}=0, \pi R$, with an $E_8$ gauge group occurring on the ten-dimensional subspace at each of the fixed points\cite{HW}. 
 In the limit that $R \rightarrow 0$ we recover the $D=10$ supergravity of the weakly-coupled heterotic $E_8 \times E_8$ theory.
  The phenomenological implications of the strongly-coupled heterotic $E_8 \times E_8$ theory may be studied by considering compactifications 
 of the $D=11$ supergravity theory, with gauge interactions confined to the 9-branes at the ends of the universe, possibly augmented 
 by 5-branes at intermediate points.  The attempts to obtain realistic phenomenology from this heterotic M-theory are reviewed in \S 4.
 
  The fact that in heterotic M-theory the gauge group lives on subspaces of the full space-time in which
  gravity propagates raises the possibility that other string theories might generate realistic models if they too admit p-branes. In fact, by considering
   the compactification of Type I open string theory on a circle of radius $R$, it is easy to see that in the 
   ``T-dual'' theory, in which the compactification is on a circle of radius $R^{-1}$ (in string units), the string satisfies Dirichlet boundary conditions, 
 in the compactified coordinate,   {\it i.e.} the ends of the string are forced to lie on fixed, eight-dimensional hyperplanes, D8-branes\cite{Polch}; 
 including Wilson lines for the $SO(32)$ gauge group, there are sixteen D8-branes on which the open strings begin and end,
  each with a massless $U(1)$ gauge boson.  In the bulk, 
 the closed superstrings are of Type IIA. Proceeding in this way with additional dimensions, we get to Type IIA superstring 
 with a collection of Dp-branes with $p$
  even, and to Type IIB with $p$ odd. If $n$ parallel Dp-branes are at the same point in their transverse space, the gauge group is enlarged to $U(n)$.
  Thus we are led to consider orientifold compactifications of Type IIB superstrings, and in \S 5 we review their success in generating realistic 
  phenomenology.
  
All of the approaches that we have described so far are ``top-down''. For example, in the models described in \S 2
 one compactifies the $E_8 \times E_8$ heterotic string on a Calabi-Yau manifold with Euler character $\chi_E = \pm 6$. With the 
 ``standard embedding'' this leads to a three-generation supersymmetric model with an $E_6$ observable gauge group, which may be broken by 
 the addition of Wilson lines. TeV scale supersymmetry breaking is assumed to arise from non-perturbative effects, such 
 as hidden-sector gaugino condensation. An alternative ``bottom-up'' approach has recently been proposed \cite{AIQU}, in which the 
 standard-model gauge group, as well as some of the matter, arises naturally from a stack of D3-branes living at an orbifold fixed point of the 
 compactified dimensions. Local cancellation of Ramond-Ramond twisted tadpole charges necessitates the addition of more D-branes, which then 
 entail further D-branes to cancel twisted tadpoles at other fixed points, until eventually a consistent theory is 
 obtained free of local and global tadpoles. The idea is to {\it start} with (a large piece) of what is needed, and to build up  from there, so that 
 central features are relatively insensitive to the global disposition of D-branes and Wilson lines. In \S 6 we describe this approach more fully,
  as well as our own attempt\cite{BKL} to get realistic phenomenology from it.

\section{Calabi-Yau compactifications of the weakly coupled heterotic string}
The first attempt to make the heterotic $E_8\times E_8$ string model 
physically realistic was the seminal paper by Candelas et al \cite{CHSW}.
In this paper the 10-dimensional Minkowski space-time, was replaced with 
a product of a  3+1-dimensional maximally symmetric space-time and some 
internal compact manifold ${\cal M}$. For the resulting 3+1-dimensional 
effective model to possess precisely ${\cal N}=1$ local supersymmetry, 
consistency requires the 3+1-dimensional space-time to be Minkowskian and 
 ${\cal M}$ to be Ricci-flat and K${\rm {\ddot a}}$hler if Riemannian.
Such six-dimensional spaces are called Calabi-Yau (CY) spaces. They have $SU(3)$ holonomy 
and vanishing first Chern class $(c_1({\cal M})=0)$.
Further restrictions arise from requiring that the effective model 
in four dimensions is anomaly free. This is guaranteed if the following 
relationship is satisfied in vacuum: 
\begin{equation}
30\int {{\rm Tr}R\wedge R}=\int {{\rm Tr}F\wedge F}
\label{anomaly}
\end{equation}
The simplest way to satisfy (\ref{anomaly}) is to set the background 
spin connection equal to the background Yang-Mills connection. Since 
the compactified space has $SU(3)$ holonomy, one seeks an SU(3) sub-bundle 
of the $E_8\times E_8$ Yang-Mills bundle. The correct numerical coefficient 
is obtained if the SU(3) is embedded entirely in one $E_8$ factor leaving 
an effective $E_6\times E_8$ gauge theory\footnote{The fibre of one 
$E_8$ factor of the Yang-Mills bundle decomposes as $248 \rightarrow (78,1)\oplus
(27,3)\oplus(\bar{27},\bar{3})\oplus (1,8)$ upon restriction to 
the $E_6\times SU(3)$ maximal subgroup.}. 
This results in standard $E_6$ families of quarks and leptons in the 27-dimensional 
 representation. The number of families is given by 
$\frac{1}{2}|\chi_E({\cal M})|$, where $\chi_E({\cal M})$ is the Euler characteristic of the 
manifold ${\cal M}$. Thus a manifold with Euler number $\pm 6$ will provide 
us with three families. Another must for low energy phenomenology is 
that the fundamental group $\pi_1({\cal M})$, which measures the ability 
to break $E_6$ to a more realistic gauge group, must be at least 
$Z_3$. Thus realistic phenomenology in this framework seems to require 
the Calabi-Yau manifold to be multiply-connected.

Let us first discuss the web of different simply-connected Calabi-Yau manifolds
and attempt to find such manifolds with the required Euler number. 
Some of these constructions will be useful in our discussions of heterotic 
M-theory.
The Euler number is given in terms of the Hodge numbers by the following 
expression
\begin{equation}
\chi_{E}({\cal M})=\sum_{r=0}^{{\rm dim}{\cal M}}(-)^{r} b_r,
\;\;\;\;b_r=\sum_{p=0}^r b_{p,r-p}.
\end{equation}
and the Hodge numbers, which are defined as the dimensions of the 
Dolbeault cohomology groups, 
$b_{p,q}:={\rm dim} H_{\bar{\partial}}^{p,q}({\cal M})$, are topological 
invariants.
There are many ways to build Calabi-Yau manifolds. In particular, CY manifolds have been constructed
 as complete intersections in products of projective spaces; intersections in which some of the additional
  factors in the embedding spaces are provided by algebraic surfaces such as $almost\;del\;Pezzo$ surfaces; 
complete intersections in 
products involving {\it almost Fano 3-folds}; embeddings in 
weighted projective spaces, and generalised {\it flag spaces}. 
A CY manifold can be defined as a complete intersection 
in a product of projective spaces 
$P_1^{n_1}\times P_2^{n_2} \times \cdots P_m^{n_m}$ subject to $K$ polynomial constraints on the coordinates;
$n_r$ is the dimension of the $r$th projective space. The total Chern class 
 is then
\begin{equation}
c({\cal M})=\frac{\prod_{r=1}^m(1+J_r)^{n_r+1}}{\prod_{a=1}^K (1+
\sum_{s=1}^m q_a^s J_s)}
\label{chern}
\end{equation}
where $q_a^r$ denotes the degree of the $a$th  defining polynomial in the coordinates of the $r$th projective space, and $J_r$ 
is the K${\rm \ddot{a}}$hler $(1,1)$ form over $P_r^{n_r}$. The Chern classes
are obtained by expanding (\ref{chern}). 
On expansion, the first Chern class is 
$
c_1=\sum_{r+1}^m(n_r+1-\sum_{a=1}^K q_a^r)J_r$. 
For a Calabi-Yau manifold
$c_1=0$ and $\sum_{a=1}^{K}q_a^r=n_r+1$ for all $r=1,\cdots,m $. 
When $\sum_{a=1}^Kq_a^r\leq n_r+1$ (with $<$ for at least one $r$), the configuration 
is {\it almost-ample}. When the inequality holds for all $r$ we talk about {\it ample 
configurations}. In two and three dimensions we say {\it (almost) del Pezzo}
and {\it (almost) Fano} instead of ``(almost) ample''.
The total Chern class for a complex vector bundle $V$ over a complex manifold 
${\cal M}$ is $c(V)=1+\sum_{r} c_r(V)$ with the Chern classes $c_r(V)\in H^{2r}({\cal M}),\;\;r=1,\cdots,min({\rm rank}V,{\rm dim}{\cal M})$. These 
represent the cohomology classes of the coefficient forms which are found in 
the expansion of the characteristic polynomial 
${\rm det}[1+\lambda {\bf F}]=1+\lambda {\rm Tr}[{\bf F}]+ 
\lambda^2({\rm Tr}[{\bf F}\wedge {\bf F}]-2[{\rm Tr}{\bf F}]^2)+\cdots$ where 
${\bf F}$ is a suitably normalized curvature two-form on $V$.
%

For realistic applications the Calabi-Yau space should be multiply connected. The simplest method of 
constructing multiply-connected varieties is to construct a variety ${\cal Y}$ 
which admits the {\it free action} of some group $G$, and then pass to the quotient space 
${\cal M}:=\{{\cal Y}/G\}$. 
Now, CY $n$-folds ($n>1$) admit no continuous 
holomorphic symmetries, so one seeks a finite group $G$ with a holomorphic and free action on ${\cal Y}$. 
The Euler number of the quotient space is given by
\begin{equation}
\chi_E({\cal M})=\frac{\chi_E({\cal Y})}{|G|}
\end{equation}
where $|G|$ denotes the order of $G$.
Freely acting symmetries are of course rather more difficult to find than 
symmetries which have fixed points sets. In the latter case, we expect singularities to occur in the quotient space ${\cal M}=\{{\cal Y}/G\}$. String theory 
can still make sense in such singular spaces (orbifolds) and phenomenologically viable models are discussed in \S 3.

The results from the complete intersection Calabi-Yau manifolds (CICY) embedded in products of projective spaces, which are prominent candidates for the 
compactification of the heterotic string are as follows. Using the $c_1({\cal M})=0$ 
condition it was recognized in \cite{PGREEN}  that all CICYs can be described 
by finitely many polynomials in products of projective spaces. Each configuration leads to a family of Calabi-Yau spaces whose generic member is smooth. By 
a computer classification, 7868 configurations with Euler numbers between \mbox{-200} and 0 were found in \cite{Candelas,Lutken}. In \cite{GOTHIC}
 all their Hodge numbers were calculated. 265 different combinations occur.  Application of a 
theorem of Wall, which states that the homotopy types of CY-threefolds 
${\cal M}$ can be classified by their Hodge numbers, their topological triple 
couplings $K_{ijk}^0 :=\int_{{\cal M}} J_i\wedge J_j \wedge J_k$ and 
$c_2.J_i:=\int_{{\cal M}}c_2 \wedge J_i$, revealed that there are at least 
2590 topologically different members in this class. 
Green and H\"{u}bsch\cite{PGREEN2} have shown that all families of CICYs are connected by the process described in
 \cite{PGREEN} of contracting a family to a nodal 
configuration and performing a small resolution of the latter. Certain quotients of them by discrete groups were 
constructed in \cite{YAU} which have Euler 
number $\chi=-6$. Two of them, have a non-trivial $Z_3$ fundamental group 
and give rise to heterotic string compactifications with three generations and 
a natural mechanism for breaking the $E_6$ group by Wilson lines. 
In \cite{YAU1} 
a 3-generation model was constructed by dividing the 
manifold 
$$\left(\begin{array}{c}
P^3 \\ P^2 \end{array}\right|
\left|\begin{array}{cc}
3 & 1  \\
0 & 3 \end{array}\right)_{-54}^{8}$$ 
by a $G=Z_3\times 
Z_3$ symmetry and resolving the singular quotient.
The second example, the Tian-Yau manifold, is the only weakly coupled 
 case that has been explored phenomenologically in some 
depth\cite{GROSS}, and is given 
by the configuration matrix
$$\left(\begin{array}{c}
P^3 \\ P^3 \end{array}\right|
\left|\begin{array}{ccc}
3 & 0 & 1 \\
0 & 3 & 1\end{array}\right)_{-18}^{14}$$ 
whose  topological Euler number $\chi_E=-18$ and 
Hodge number is $b_{1,1}=14$. 
The integers denote the degrees of the defining system of polynomials of 
constraints for the embedded 3-fold ${\cal M}$.
Thus the above Tian-Yau manifold represents the system of polynomials of bi-degree (3,0),(0,3) and (1,1) in the homogeneous coordinates $x_{i}, y_{i}
\quad (i=0,1,2,3)$ of $P^3\times P^3$ \cite{GROSS}
\begin{eqnarray}
\sum x_{i}^3+a_3 x_0 x_1 x_2+a_2 x_0 x_1 x_3 +a_1 x_0 x_2 x_3 +a_0 x_1 x_2 x_3&=&0 \\ \nonumber 
\sum y_{i}^3+b_3 y_0 y_1 y_2  +b_2 y_0 y_1 y_3+b_1 y_0 y_2 y_3 +b_0 y_1 y_2 y_3&=&0 \\ \nonumber
\sum c_{ij}x_{i}y_{j}=0, \qquad (c_{00}&=&1)
\end{eqnarray}
where $a_{i}, b_i ,c_{ij}$ are the 23 
independent complex parameters that represent deformations of the complex 
structure of the  Tian-Yau manifold, and whose number equals the Hodge number $b_{2,1}=23$.
The quotient of this manifold by the $Z_3$ group acting by
\begin{equation}
(x_0,x_1,x_2,x_3,y_0,y_1,y_2,y_3)\mapsto (x_0,x_1,x_2,\alpha x_3,y_0, 
y_1,y_2,\alpha^2 y_3),\;(\alpha=e^{\frac{2\pi i}{3}})
\end{equation}
on the homogeneous coordinates
 yields a simple realization 
of a three-generation compactification, which is diffeomorphically equivalent 
to the one in the first example.
Indeed, since $|Z_3|=3$, the Euler number of the quotient manifold is $\chi_E=
-18/3=-6$. The rest of the topological data are determined as follows. Of the 
23 $(2,1)$ forms of the Tian-Yau manifold 9 transform trivially under $Z_3$ 
and these descend to harmonic forms of the quotient manifold \cite{GROSS}. 
Consequently, $b_{2,1}=9$ and therefore, since $\frac{1}{2}\chi_E({\cal M})=-3=
b_{1,1}-b_{2,1}$, we have $b_{1,1}=6$.
In \cite{GROSS} it was also shown how a combination of symmetry breaking by 
Wilson lines and Higgs scalar VEVs can lead to low-energy  models with properties close to the minimal supersymmetric standard model. 
More specifically, an embedding of the fundamental group $\pi_1({\cal M})=Z_3$ of the 
quotient CY manifold in $E_6$  results in an $SU(3)\times SU(3) \times SU(3)$ 
gauge group. Further, intermediate scale (IS) gauge symmetry breaking  
 by scalar VEVs breaks the symmetry down to the standard model gauge group, with three 
generations of quarks and leptons remaining massless before electroweak breaking.
Discrete symmetries in the model are important for eliminating dimension-four
 baryon- and lepton-number violating operators. However, baryon and lepton 
number could be violated by higher-dimensional operators which are suppressed 
by inverse powers of the intermediate scale. Experimental bounds on nucleon decay require that 
$M_{IS}>O(10^{14})$GeV.

Both models have (2,2) world-sheet supersymmetry which naturally yields an 
$E_6$ effective GUT group. On the other hand, as  has been emphasized by 
Greene, Distler and Kachru\cite{GREENEb,Distler,Kachru}, (0,2) Calabi-Yau models provide a much broader class 
of compactifications in which one can also naturally obtain $SO(10)$ or 
$SU(5)$ as the effective gauge group. In fact, Kachru constructed  four examples in this class of models which yield 
three generations of chiral fermions in the normal unification representations of $E_6$ and $SU(5)$. Two of the examples, 
with $SU(5)$ gauge group, are on manifolds with a non-trivial fundamental group. Furthermore, using Wilson lines as 
a tool for gauge symmetry breaking, the simple gauge 
group is broken down to $SU(3)\times SU(2) \times U(1)$. 
When compactifying on a manifold ${\cal M}$ with $\pi_1({\cal M})
\neq 0$, expectation values may be given to Wilson lines around the noncontractible loops $\gamma$ in ${\cal M}$ 
\begin{equation}
U_{\gamma}=P e^{(\oint_{\gamma}Adx)}
\end{equation}
This amounts to a choice of a homomorphism from $\pi_1({\cal M})\rightarrow 
G$, where $G$ is the space-time gauge group. This leaves with a vacuum with 
$G$ broken to the subgroup of $G$ which commutes with $U_{\gamma}$.
The data which enters in specifying a (0,2) Calabi-Yau model is a choice of 
the Calabi-Yau manifold ${\cal M}$ and stable, holomorphic vector bundles $V_1$ and $V_2$ (representing the vacuum configurations of the gauge fields 
in the observable and hidden $E_8$ of the heterotic string) satisfying 
\begin{eqnarray}
c_2{(\cal M)}&=&c_2(V_1)+c_2(V_2) \\
c_1(V_{1,2})&=&0 \bmod 2.
\end{eqnarray}
Here the $c_i$ are the Chern classes of the vector bundles in question.
The first equation is the well-known anomaly cancellation condition, while the 
second is the requirement that $V_1$ and $V_2$ admit spinors.

\section{Orbifold compactifications of the weakly coupled heterotic string}
Geometric models in which calculations can be more easily and more fully
 carried out can be obtained from the heterotic string by replacing the 
 Calabi-Yau manifold by a (toroidal) orbifold. 
 (For a review see Bailin and Love\cite{BLPhysRep}.) A toroidal orbifold is the closest we 
 can get to a torus while reducing the supersymmetry from $\mathcal{N}=4$ to
  $\mathcal{N}=1$. To go from an underlying torus to an orbifold, points 
  on the torus $T^6$ are identified when they are linked by the action of an 
  element of a discrete group $P$, called the point group, which acts 
  crystallographically on the lattice defining equivalent points on the torus. 
  Thus, the orbifold is the space $T^6/P$. Provided that $P$ is a discrete 
  subgroup of $SU(3)$, the surviving supersymmetry is  $\mathcal{N}=1$. This 
  requires $P$ to be specific $Z _N$ or $Z _M \times Z _N$ groups.
  
  The massless states for an orbifold live either in the untwisted sector,
   where the world sheet boundary conditions are the same as for the underlying 
   torus, or in a twisted sector, with boundary conditions twisted by an element
    of the point group ($\theta ^k, \quad k=1,2,...,N-1$ in 
    the case of a $Z _N$ pointgroup.) In a twisted sector the bosonic degrees
     of freedom of the string return to their original values only up to the 
     action of a point group element in going around a closed string.
     
     Focussing on $Z _N$ orbifolds, the $\theta ^k$ twisted sector is in
      fact a set of twisted sectors located at the fixed points of  $\theta ^k$,
      as we now discuss. The twisted boundary condition on the bosonic degrees of 
freedom of the heterotic string implies, in particular, that the centre-of-mass coordinate ${\bf Z }$
 of the string, in a complex basis ($Z^1\equiv \frac{1}{\sqrt 2}(X^1+iX^2)$ etc.) is 
at an orbifold fixed point, i.e.
\begin{equation}
(\theta ^k, {\bf \ell}){\bf Z}=\theta ^k {\bf Z}+{\bf \ell}={\bf Z}
\label{FP}
\end{equation}
for some lattice vector ${\bf \ell}$. There is a distinct string sector for each fixed point.
 
In general, quark and lepton generations can arise from the untwisted sector or from the 
various twisted sectors at various fixed points. In the first instance, the number of 
generations turns out to be larger than three. However, this number can be  adjusted by 
introducing Wilson lines, {\it i.e.} non-zero quantities $U$ of the form
\begin{equation}
U \sim \oint A_m \ dx^m
\label{Wilson}
\end{equation}
where $m$ runs over compact ``manifold'' coordinates, the integral is around some 
closed loop on the underlying torus not contractible to zero, and $A_m$ are the 
components of some ten-dimensional gauge field with zero field strength. These 
quantities can be gauged away by means of non-single-valued gauge transformations. 
In this alternative  formulation, the Wilson lines (\ref{Wilson}) are no longer present but instead,
 the bosonic degrees of freedom associated with $E_8 \times E^{\prime} _8$ acquire extra shifts 
 upon a circuit of the torus. As a consequence, the boundary conditions  for the twisted sectors 
 of the orbifold are modified. These extra shifts and this modification depend 
 on the particular fixed point sector. For example, for the case of a $Z_3$ orbifold, if the 
 embedding of the point group element $\theta$ in the gauge group is represented by the shift $\pi v^I$ 
  on the boundary conditions of the bosonic degrees of freedom, and the embedding of the lattice basis 
  vector ${\bf e}_{\rho}$ with which the Wilson line is associated  is represented by the shift $\pi a^I_{\rho}$, 
  then the left-mover mass formula is modified to
\begin{equation}
\frac{1}{2} m^2_L=\frac{1}{2} \sum_{I=0}^{16}\left( p^I_L+v^I+r_{\rho} a_{\rho}^I \right)^2 +\tilde{N} -\frac{2}{3}
\label{ML}
\end{equation}  
with $\tilde{N}$ the oscillator term, and $r_{\rho}=0,\pm1$ for the various fixed points of $\theta$.
 The spectrum of massless states is now different for different fixed points because of the influence of
  the embeddings $ a_{\rho}^I$ of the lattice basis vector ${\bf e}_{\rho}$ (the Wilson lines.)  
  In this way, it turns out to be possible to adjust the number of generations to three by a suitable choice 
  of Wilson lines\cite{3genwilson}. Wilson lines also allow the gauge group to be adjusted by 
  generalised GSO projections on the untwisted sector.
  
  The Yukawa couplings for orbifold compactifications of the heterotic string are subject to a number of 
  useful selection rules. For example, for a non-zero three-point function, the product of the (three) space-group 
  elements for the fixed points with which the states are associated should contain the identity element of the 
   space group. In particular, the product of the point group elements for the three twisted sectors should be the 
   identity. Also, if we bosonize the NSR right-moving fermionic degrees of freedom so that the  right 
   movers are represented by a five-component momentum (H-momentum), then H-momentum should be conserved.
   
   Indeed, the Yukawa couplings are fully calculable using conformal field theory methods. For example, 
   for the case of the $Z_3$ orbifold with point group generated by $\theta$, and for large values of the torus 
   radii $R_{2i-1}, \ i=1,2,3$ for the three complex planes, the Yukawa coupling $Y$ between 
   three $\theta$-twisted sectors takes the form\cite{3genwilson}:
   \begin{equation}
   \begin{array}{rcl}
   Y & \sim & 1 \qquad {\rm for } \quad d_{2i-1}=0  \nonumber \\
   & \sim & \exp\left( -\frac{1}{2\pi \sqrt{3}} \sum_i R^2_{2i-1} \right) {\rm for} \quad d_{2i-1}=\pm1,\pm2 \label{Yuk} 
   \end{array}
   \end{equation}
In (\ref{Yuk}) 
\begin{equation}
d_{2i-1}=p^1_{2i-1}-p^2_{2i-1}
\label{dp}
\end{equation}
where $p^J_{2i-1}$ characterizes the $\theta$-twisted sector fixed point for the $J$th state, and
 $p^J_{2i-1}=0,\pm 1$ for each value of $i$. The sizes of the Yukawas are controlled by the ``distances'' 
 $d_{2i-1}$ between the fixed points in the three complex planes, and the exponential dependence of the 
 Yukawas on $R^2_{2i-1}$ allows for mass hierarchies to develop when the Higgs expectation values are turned on.
 
 A problem with orbifold compactifications of the weakly coupled heterotic string is that the (MSSM)
  renormalization group running of the observed gauge coupling constants requires unification to take place\cite{mssm1016} at 
  an energy scale of order $10^{16} \ {\rm GeV}$. However, the tree level heterotic string theory requires unification 
  of gauge coupling constants to occur at the string scale\cite{mstring} which is fixed to be about $10^{18} \ {\rm GeV}$.
   It is possible to overcome this difficulty by taking account of the moduli-dependent corrections to the 
   gauge kinetic function, but only for values of the moduli an order of magnitude larger than those obtained by minimizing 
   the effective potential\cite{largeT} (though this minimization may itself be suspect because of the dilaton stabilization problem.)
    The problem may also be overcome by adding (vector-like) matter at a scale below $10^{16} \ {\rm GeV}$.
    
 \section{Heterotic M-theory}


M-theory on the orbifold $S^1/Z_2$ is believed to describe the strong coupling limit of
 the $E_8 \times E_8$ heterotic string \cite{HW}, and thus, constitutes a particularly
  interesting starting point for particle phenonemonology.
At low energy, this theory is described by 11-dimensional supergravity coupled 
to two 10-dimensional $E_8$ gauge multiplets residing on the two fixed points of the orbifold.
 Compactifications leading to ${\cal N}=1$ supersymmetry in four dimensions are based on space-times of the
  structure  $M_{11}=S^1/Z_2 \times {\cal M} \times M_4$, where ${\cal M}$ is a CY-threefold and $M_4$ is
   flat Minkowski space. Heterotic M-theory vacua with both standard and non-standard embeddings 
\cite{STEVE} have 
been studied.

A novel construction of a CY threefold ${\cal M}$, which 
has been exploited in heterotic M-theory, is one in 
which the manifold can be represented as an elliptic fibration over a surface 
${\cal S}$. 
The elliptically fibred CY manifolds including 5-branes are the direct descendents 
of those those that arise in the weakly coupled case which were discussed in \S 2. 
In particular, Donagi {\it et al}\cite{DONAGI} constructed elliptically fibred 
CY spaces, which lead to non-perturbative ${\cal N}=1$ supersymmetric vacua  
of heterotic M-theory, satisfying the phenomenological requirement that the 
low energy gauge group is a grand unified gauge group such as $SU(5), \quad SO(10)$ 
or $E_6$ with three generations of chiral fermions. 
To do so the background gauge fields must lie in an $SU(N)$ subgroup 
of $E_8$ and there is a constraint on the third Chern class of the gauge 
bundles. The form of this constraint for elliptically fibred CY 
three-folds was first presented in \cite{CURIO,Andreas,FMW}. One then chooses a 
particular Calabi-Yau threefold and finds the relevant $SU(N)$ bundle 
which is a solution of the three-family condition. That this be a 
true $SU(N)$ bundle provides  a second constraint. They also allowed 
the presence of NS 5-branes in the vacua. The requirement that the 
vacua be anomaly free and ${\cal N}=1$ supersymmetric leads to a cohomology 
condition that fixes the homology class of the 5-branes in terms of Chern 
classes of the gauge bundles and the CY tangent space.
The five-branes carry unitary groups on their world volumes. Enhancement 
of the gauge group occurs when either two or more five branes 
overlap, or a single 5-brane degenerates in such a way such that two parts 
of the wrapping curve come close together in the CY manifold. They can be considered as a further set of hidden sectors.

More precisely, an elliptically fibred Calabi-Yau threefold
${\cal M}$ consists of a base ${\cal S}$, which is a complex 
two-surface, and an analytic map
\begin{equation}
\pi:{\cal M}\rightarrow {\cal S}
\end{equation}
with the property that for a generic point $b\in {\cal S}$, the fibre 
\begin{equation}
E_b=\pi^{-1}(b)
\end{equation} is an $elliptic\;curve$, that is a Riemann surface 
of genus one. In addition, it is usually required that there exists a global 
section, denoted $\sigma$, defined to be an analytic map
\begin{equation}
\sigma:{\cal S}\rightarrow {\cal M}
\end{equation}
that assigns to every point $b\in {\cal S}$ the zero element 
($\sigma{b}={\cal O}$) of the elliptic curve $E_b$ group \footnote{One can define a homomorphism 
$f$ from the additive group of complex numbers onto the group of complex 
points on a cubic algebraic curve via elliptic functions. The kernel of this 
homomorphism is a lattice $L$. The factor group, $\frac{C}{L}=\frac{C}{{\rm Ker}(f)}\sim f(u),u\in C$, of the complex plane modulo the lattice $L$ is 
isomorphic to the group of complex points on the elliptic curve. Therefore, 
the group of complex points on the elliptic curve is that of a torus, the direct product 
of two cyclic groups.} We recall some of the properties of an elliptic curve. The defining relation for a general projective cubic is provided by 
the Weierstra$\ss$ equation
\begin{eqnarray}
P(x,y,z)&=&4(x-e_1 z)(x-e_2 z) (x-e_3 z)-y^2 z \\
&=&4 x^3-g_2 x z^2-g_3z^3-y^2 z=0, \;\;g_2,g_3 \in C
\end{eqnarray}
where  $e_i, \; (i=1,2,3)$ are the three roots, of the cubic $4 x^3-g_2 x-g_3$. The curve is smooth. That is, it fits into the 
projective space $P^2$ without singularities, which is to say that the gradient of the defining relation $P=0$ does not vanish any place on the curve. 
The invariants of the cubic ($g_2$ and $g_3$) encode the different complex 
structures one can put on the torus. The discriminant $\Delta$ of the curve is given by
\begin{equation}
\Delta=g_2^3-27 g_3^2 
\end{equation}
The curve has genus one if it is non-singular, {\it i.e.} $\Delta \neq 0$.

Using the spectral cover formalism,  Donagi {\it et al}\cite{DONAGI} 
constructed semi-stable
holomorphic gauge bundles with fibre groups $G_1$ and $G_2$ over the 
orbifold fixed planes of heterotic M-theory. By restricting the structure groups 
to be 
\begin{equation}
G_i=U(N_i) \; {\rm or} \; SU(N_i)
\end{equation}
they showed that the base ${\cal S}$ is restricted to almost del Pezzo, 
Hirzebruch or Enrique surfaces, as well as blow-ups of Hirzebruch surfaces.
Del Pezzo surfaces have a positive first Chern class ($c_1({\cal S})>0$).
Some examples of Hirzebruch and almost del Pezzo surfaces are given by the 
configurations
\begin{equation}
F_1=\left(\begin{array}{c}
P^2 \\ P^1 \end{array}\right|
\left|\begin{array}{c}
1  \\
1 \end{array}\right)_{4},\;\; 
 \left(\begin{array}{c}
P^2 \\ P^1 \end{array}\right|
\left|\begin{array}{c}
3  \\
1 \end{array}\right)_{12} etc.
\label{delpezzo}
\end{equation}
where subscripts denote Euler characteristics.
Donagi {\it et al}\cite{DONAGI} also showed that the 
second Chern class $c_2({\cal T}_{\cal M})$ of the holomorphic 
tangent bundle of ${\cal M}$ is given by 
\begin{equation}
c_2({\cal T}_{\cal M} )=c_2({\cal S})+11 c_1({\cal S})^2+12 \sigma c_1({\cal S})
\end{equation}
 For the allowed cases discussed above  the first and second Chern 
classes of the base ${\cal S}$ are known.
The ingredients of the spectral cover construction of semi-stable holomorphic 
gauge bundles include the Chern classes of an $SU(N)$ gauge bundle $V$, 
which are\cite{CURIO,Andreas,FMW}
\begin{eqnarray}
c_1(V)&=&0 \\
c_2(V)&=&\eta\sigma-\frac{1}{24}c_1({\cal S})^2 (N^3-N)+\frac{1}{2}(\lambda^2-
\frac{1}{4})N\eta \left( \eta-Nc_1 ({\cal S}) \right) \\
c_3(V)&=&2 \lambda \sigma \eta(\eta-N c_1({\cal S}) 
\end{eqnarray}
The $SU(N)$ gauge bundle is determined by two 
line bundles ${\cal X}$ and ${\cal N}$, and $c_1({\cal X}) \equiv \eta $. 
Further, the requirement that the first Chern number $c_1({\cal N})$
 is integral 
requires\cite{DONAGI} that the parameter $\lambda=m+\frac{1}{2}$ with $m \in Z$, for $N {\rm odd}$, or $\lambda=m$ and $\eta=c_1({\cal S}) \bmod 2$  for $N$ even.

The third ingredient for defining a non-perturbative vacuum 
is  a set of 5-branes wrapped on holomorphic curves within ${\cal M}$. The cancellation of anomalies leads to the condition
\begin{equation}
[W]=c_2({\cal M})-c_2(V_1)-c_2(V_2)
\end{equation}
where the class $[W]=\sum_{n=1}^K[J^{(n)}]$ must be $effective$ \cite{DONAGI} 
and $J^{(n)}$ are sources arising from $K$ 5-branes located in the orbifold 
interval.
The number of generations is given by
\begin{equation}
N_{gen}=\frac{1}{2}\int_{{\cal M}}c_3(V)=\lambda\eta(\eta-N c_1({\cal S}))
\end{equation}
The three-family condition becomes
\begin{equation}
3=\lambda\eta(\eta-N c_1({\cal S})=\lambda[W_{\cal S}^2-(24-N) 
W_{\cal S} c_1({\cal S})+12(12-N)c_1({\cal S})^2]
\label{tria}
\end{equation}
where the second form of the condition is expressed in terms of the effective 
class $W_{{\cal S}}$.
If $G$ denotes the structure group and $H$ its commutant subgroup, Donagi {\it et al}\cite{DONAGI} concentrated on the 
cases where $H=E_6, \quad SO(10), {\rm or} \quad SU(5)$. For instance, one 
can obtain an  observable gauge symmetry $H=SU(5)$ with the base ${\cal S}$ 
described by the del Pezzo surface $dP_9$ with configuration matrix the 
second matrix in eqn (\ref{delpezzo}). In this case $c_2({\cal S})=
\chi_E{({\cal S})}=12$. Also $N=5$ and $\lambda=\frac{3}{2}$ by choosing $m=1$. By choosing
 the component of the five-brane class, $W_{{\cal S}}$ in $dP_9$ appropriately, 
the three family condition (\ref{tria}) is satisfied. Thus this is an 
example of a three-generation model with $SU(5)$ GUT symmetry and five-branes present. In \cite{DONAGI} it was also shown that 
elliptically fibred CY threefolds with an Enriques 
base never admit an effective five-brane curve if one requires that there be 
three families.      

Since the theories described in this framework are GUTs, it is important 
to know whether it is  possible to break the GUT symmetry spontaneously  down to 
the standard model. In \cite{DONAGI} it was demonstrated that this is
indeed possible using Wilson lines.

\section{Type II orientifold models}
An alternative approach to the strongly coupled heterotic string (M-theory) is to stick with weakly 
coupled string theory, but to study a type IIB string theory with D-branes. In particular, Type IIB 
theories compactified on an orientifold have been considered. In the simplest case, a type IIB string
compactified on an orbifold is also quotiented by the world sheet parity operation $\Omega$, {\it i.e.} 
the space-like world sheet coordinates $\sigma$ and $-\sigma$ are identified. The theory then contains open 
strings as well as closed strings, with the open strings beginning and ending on Dp-branes (solitonic 
solutions with a $p+1$-dimensional world volume.) Such theories are equivalent to orbifold-compactified 
$SO(32)$ type I theories. The presence of some Dp-branes is required for consistency of the 
theory because of the need to cancel (divergent) tadpoles coming from amplitudes on Klein-bottle
 world sheets. For example, in the absence of anti-Dp-branes, for $Z_N$ orientifolds with $N$ odd,
  only D9-branes are allowed and 32 D9-branes are needed to cancel (untwisted) tadpoles. For 
  $Z_N$ orientifolds with $N$ even, D5-branes may also be present. More generally, anti-Dp-branes 
  may be present and D5-branes and anti-D5-branes may be present when $N$ is odd. (There is a T-dual picture
   in which D9-branes are replaced by D7-branes, and D5-branes by D3-branes.)
   
   Models are characterized by the action of the $Z_N$ point group on the Chan-Paton matrices.
    This action is constrained by the need to cancel also tadpoles involving fields from the twisted 
    sectors arising from amplitudes on cylinder and M\"{o}bius strip world sheets, as well 
    as Klein bottle world sheets. If the Chan-Paton matrices for an open string with end points
     on a Dp-brane and a Dq-brane are denoted by $\lambda^{pq}$, then the action of an 
     orbifold group element $g$ on the Chan-Paton matrix is of the form
\begin{equation}
g: \quad \lambda^{pq} \rightarrow \gamma_{g,p} \lambda^{pq}  \gamma^{-1}_{g,q}
\label{gamg}
\end{equation}
where   $\gamma_{g,p}$ and $ \gamma_{g,q}$ are the unitary matrices. Wilson lines may be introduced, 
much as for orbifolds, with a similar action on the Chan-Paton matrices. Similarly for the action 
of $\Omega$. 

In orientifold models, gauge  fields arise from open strings (with Chan-Paton factors) beginning and 
ending on a set of Dp-branes. In the most promising models to date\cite{promise}, the observable gauge 
group occurs in the $99$ sector (where the $pq$ sector denotes open strings 
beginning and ending on Dp-branes and Dq-branes) with chiral massless matter occurring in $99,55$ and $59$
 sectors. Allowed states are required to be invariant under the combined action of the orbifold group 
 on both the explicit oscillator factors in the state and the Chan-Paton matrix. To avoid
  extended supersymmetry the Dp-branes are located at orbifold singularities (fixed points). 

Three generations (together with some extra matter in vector-like representations) arise in a natural way.
 First, matter in bi-fundamental representations of gauge group factors occurs in the $99$ sector so that 
 a $U(3) \times U(2)$ factor provides quark doublets in the $({\bf 3}, {\bf \bar{2}})$ representation of 
$SU(3) \times SU(2)$. Also, the bosonic component of a massless matter supermultiplet in the $99$ sector is 
a state of the form
\begin{equation}
\psi ^i_{-\frac{1}{2}} |0 \rangle \lambda^{99 (i)}_{ab} \qquad i=1,2,3
\label{state}
\end{equation} 
where $\psi ^i_{-\frac{1}{2}}|0 \rangle $ is the oscilltor factor, $\lambda^{99 (i)}_{ab}$ is the Chan-Paton factor, and 
$i=1,2,3$ refers to the three complex planes for the compact string degrees of freedom 
($Z^1\equiv \frac{1}{\sqrt 2}(X^1+iX^2)$ etc.)  The weight vectors $\rho ^i$ for these matter states 
are subject to the projections
\begin{equation}
\rho ^i.V_{\theta}=v^i_{\theta} \bmod 1, \quad i=1,2,3
\label{weight}
\end{equation}
where $V_{\theta}$ is a shift on the $SO(32)$ lattice representing $\gamma_{\theta ,9}$, and $v^i_{\theta}$
 are the twists for the point group element $\theta$. For the case of a  $Z_3$ orientifold
 \begin{equation}
 v_{\theta}=\left( \frac{1}{3},\frac{1}{3},-\frac{2}{3} \right)
 \label{Z3}
 \end{equation}
and, since all three twists are $\frac{1}{3} \bmod 1$, we get identical projections for all three states. 
Moreover, if any Wilson line present is represented by the shift $W$, there is also a projection
\begin{equation}
\rho ^i.W=0 \bmod 1
\label{Wproj}
\end{equation}
which does not distinguish $i=1,2,3$. Thus, in the case of a $Z_3$ orientifold, the $99$ matter states and, 
in particular, the  quark doublets come in three copies. Essentially, there are the three generations 
because the compact string degrees of freedom form three complex planes.

The most realistic orientifold models constructed to date have contained anti-D-branes.
 In that case, supersymmetry is broken explicitly in an anti-D-brane hidden sector, 
 with supersymmetry breaking being transmitted gravitationally to the observable sector.  
Since we expect the scale of supersymmetry breaking in the anti-D-brane sector to be the string scale 
$m_s$, the mass of the sparticles in the observable sector may be expected to be of order $m_s^2/m_P$, 
where $m_P$ is the four-dimensional Planck mass. For sparticle masses of order  $1 {\rm TeV}$ we have 
$m_s \sim 10^{11} {\rm GeV}$. Then, unification of gauge coupling constants at a scale of order 
$10^{11} {\rm GeV}$ is to be expected (modulo some subtleties to do with Kaluza-Klein 
modes and winding modes\cite{KKmodes}).  Interestingly, orientifold models can contain the extra matter
 required to allow the renormalization group equations to run to unification at this lower scale, though 
with an unconventional value for the coupling constant ratio $g_1^2/g_3^2$ which arises from the identification of the weak hypercharge 
with a specific non-anomalous $U(1)$. The string scale is necessarily of order the Planck mass in weakly 
coupled heterotic  theories. However, in type II orientifold theories with isotropic compactification on 
a scale $R=m_c^{-1}$ and gauge fields for the observable sector arising from D9-brane open strings, we have 
the relationship
\begin{equation}
\frac{m_c^3}{m_s^2}=\frac{\alpha _X m_P}{2 \sqrt2 }
\label{mc}
\end{equation}
where $\alpha _X \simeq \frac{1}{24}$. Then, the string scale can be adjusted to an intermediate scale by 
adjusting $m_c$.

There is a potential problem of too rapid proton decay with such a low string scale. However, type IIB 
orientifold theories differ from weakly coupled heterotic theories in a vital respect. Whereas
 in the latter case anomalous $U(1)$ factors in the gauge group disappear completely after the anomaly 
 has been cancelled by a Green-Schwarz mechanism, in the former case anomalous $U(1)$ gauge symmetries are
able to survive as global symmetries in the low energy theory. This different behaviour is a consequence of 
the fact that in the heterotic case 
the Fayet-Iliopouls term for the anomalous $U(1)$  involves the dilaton field whose expectation
 value is controlled by the strength of the gauge coupling constant, whereas in the type IIB orientifold case it
  involves orbifold blowing-up modes whose expectation values can approach zero as the
   orbifold limit is approached. These global $U(1)$ symmetries may forbid
    dangerous terms in the superpotential. They are also a  promising way of obtaining a quark-lepton mass
     matrix with appropriate hierarchies.
     
\section{Three-generation Type IIB orbifold model}
We noted in the \S 1 that by considering toroidal compactifications of Type I superstrings we are led naturally to Type II 
strings with D-branes. It is well known that Type IIB theory quotiented with the $Z_2$ generated by the world sheet parity operator $\Omega$ is just 
Type I theory. Thus orbifold compactifications of Type I are orientifold compactifications of Type IIB, as observed in the previous section. 
However, if we consider only orbifold compactifications of Type IIB and their T-duals we will never encounter open strings or D-branes, but there is no reason 
why we should not {\it start} from Type IIB with D-branes, and this is the basis of the ``bottom-up'' approach\cite{AIQU}. 

It is convenient, and more transparent, to work in the T-dual picture to that used in \S 5. The starting point is a stack of 
six D3-branes at a $Z_3$ orbifold fixed point that without loss of generality may be taken to be the origin $O$. Then
 the gauge vector bosons 
(and gauginos) of $U(6)$ arise from open strings that begin and end on the D3-branes. The gauge group is broken to $U(3) \times U(2) \times U(1)$
 when the action of the
 generator $\theta$ of the $Z_3$ point group is embedded in the Chan-Paton indices by the matrix 
 \begin{equation}
 \gamma_{\theta, 3}={\rm diag}(I_3,\alpha I_2,\alpha^2)
 \end{equation} 
 where $\alpha \equiv e^{2\pi i/3}$. Each of the $U(n)$ groups has its own $U(1)$ charge $Q_n$, and the only non-anomalous combination is proportional to
 (the weak hypercharge) 
 \begin{equation}-Y \equiv \frac{1}{3}Q_3+\frac{1}{2}Q_2+Q_1
 \end{equation}
  It follows that at the unification scale the three standard model gauge 
 couplings satisfy 
 \begin{equation}
 \alpha_3(m_X)=\alpha_2(m_X)=\frac{11}{3}\alpha_1(m_X)
 \label{alfax}
 \end{equation}
  corresponding to 
  \begin{equation}
  \sin^2 \theta_W(m_X)= \frac{3}{14}=0.214
  \end{equation} 
 The $33$ sector also produces $n_G=3$ generations of (supersymmetric) chiral matter in the (bi-findamental)
  representations 
 $({\bf 3},{\bf 2})_{1/6}+({\bf 2},{\bf 1})_{1/2}+({\bf 1},{\bar{\bf 3}})_{-2/3}$ of the gauge group, corresponding to $Q_L+H_u+u^c_L$.
  It is easy to see that this has non-abelian anomaly, which derives from the existence of uncancelled Ramond-Ramond (RR) twisted tadpoles at $O$. To cancel 
  them D7-branes passing through $O$ must be introduced. We choose to introduce only D$7_3$-branes with $z_3=0$, wrapping the $z_{1,2}$ complex planes, with point-group 
  embedding 
  \begin{equation}
  \gamma_{\theta, 7_3}={\rm diag}(I_{u},\alpha I_{u+3},\alpha^2 I_{u+6})
  \end{equation}
   so that the twisted-tadpole cancellation condition 
  \begin{equation}
  3 {\rm Tr}\gamma_{\theta, 3}-\gamma_{\theta, 7_1}-\gamma_{\theta, 7_2}+\gamma_{\theta, 7_3}=0
  \end{equation}
   is satisfied (for any integer $u \geq 0$). The D7-branes generate their own gauge group, and the  
  $37_3+7_33$ states complete the $n_G=3$ chiral generations together with extra vector-like matter. In all we get
  \begin{eqnarray}
3(Q_L+u_L^c+d_L^c+L+H_u+H_d+e_L^c) + \nonumber \\
(u+3)(d_L^c+\bar{d}_L^c)+u(H_u+H_d) + u(e_L^c+\bar{e}_L^c)  \label{matter}
\end{eqnarray}
The introduction of the D7-branes leads to uncancelled twisted tadpoles at the eight other orbifold fixed points with $z_3=0$, namely at $(p,\pm 1,0)$
where $p=\pm 1,0$, and at
 ($\pm 1,0,0)$.  Introducing a suitably chosen Wilson line in the $z_2$-direction cancels the RR charge  at the six points with
  $z_2 \neq 0$, and at the same time reduces the size of the D7-brane gauge group; we take $u=3\tilde{u}$ and the D7-brane gauge group is reduced to
  $[U(\tilde{u}) \times U(1+\tilde{u}) \times U(2+\tilde{u})]^3$.
   The cancellation at the remaining two fixed points is easily arranged,
   for example by introducing a stack of six D3-branes at each of them, with $\theta$ embedded as at $O$. The important point is
   that so far there are no branes, and so no twisted RR charge, at any of the fixed points in the planes $z_3= \pm 1$. Thus 
   we are free finally to cancel the untwisted tadpoles by putting an appropriate array of anti-branes  in (say) the $z_3=1$ plane. Precisely what array 
   is used does not matter. The essential point is that it {\it can} be done,  and in a way that ensures cancellation of all twisted 
   RR charges in the $z_3=1$ plane.  The anti-branes are non-supersymmetric, but this lack of supersymmetry is transmitted to the observable 
   sector (in the plane $z_3=0$) only gravitationally, by closed string states. Thus, as in \S 5, requiring that the sparticles have TeV-scale
    masses requires a string scale ($m_s$)
     of order $10^{11}$GeV. The question, therefore, is whether the 
    matter content we have derived leads to unification at a scale ($m_X$) that is consistent with such a string scale.
    
     The (one-loop) renormalization group equations give
\begin{eqnarray}
\lefteqn{\sin^2 \theta_W(m_Z) = \frac{3}{14} +} \nonumber \\
& &+ \frac{11b_2-3b_1}{3b_1+3b_2-14b_3}\Bigl[
\frac{3}{14}-\frac{\alpha_{em}(m_Z)}{\alpha_3(m_Z)}\Bigr] 
\end{eqnarray}
where $b_i \ (i=1,2,3)$ are the beta function coefficients. The matter content (\ref{matter}) gives $(11b_2-3b_1)/(3b_1+3b_2-14b_3)<0$, 
so that $\sin^2 \theta_W(m_Z)<0.214$, whereas
\begin{equation}
\sin^2 \theta_W(m_Z)|_{exp}=0.231
\label{s2w}
\end{equation}
However, if we assume that we are able to turn on string-scale 
VEVs for the D7-brane matter fields, then, since the $37_3+7_33$ matter fields are coupled to the  $7_37_3$ matter fields, we are free to give 
string-scale masses to some or all of the vector-like matter. We therefore consider the possibility of giving masses to $\alpha$ copies of 
$e_L^c+\bar{e}_L^c$, $\beta$ copies of $H_u+H_d$, and $\gamma$ copies of $d_L^c+\bar{d}_L^c$ on a scale larger than $m_X$. 
 Then using the Particle Data Group values $\alpha_{em}^{-1}(m_Z)=128.9$ and $\alpha_3(m_Z)=0.119$, the above renormalization group equation 
 gives 
 \begin{equation}
 \sin^2 \theta_W(m_Z)=0.2275 \quad {\rm and} \quad m_X=1.3 \times 10^{10}{\rm GeV}
 \label{ZX}
 \end{equation}
 when $\alpha - \beta =2$ and $\gamma - \beta =4$. 
  We can obtain a larger unification scale, and exactly the measured value (\ref{s2w}) of 
 $\sin^2 \theta_W(m_Z)$, if we assume in addition that some other $77$ sector scalars give masses to all of the extra matter, over and above that 
 in the MSSM, on a scale $m_Y$ not very much larger than $m_Z$. Then with  $\alpha - \beta =3$ and $\gamma - \beta =3$ we get  
 \begin{equation}
 \frac{m_Y}{m_Z}=13.9 \quad {\rm and} \quad m_X=1.1 \times 10^{12}{\rm GeV}
 \label{ZYX}
 \end{equation}
 Assuming an isotropic compactification, the compactification scale $m_c$ is given by\cite{KKmodes}
 \begin{equation}
 \frac{m_s^4}{m_c^3}=\alpha_D\frac{m_P}{\sqrt 2}
 \label{msc}
 \end{equation} 
 where $\alpha_D$ is the value of the unified coupling (\ref{alfax}) on the D3-branes. Open strings beginning and ending on D3-branes 
 have no Kaluza-Klein (momentum) modes, only winding modes with mass a multiple of  $m_w=m_s^2/m_c$. Thus
 \begin{equation}
 \frac{m_w}{m_s}=\left( \frac{\alpha_D\ m_P}{\sqrt 2 m_s} \right)^{1/3}
 \label{mw}
 \end{equation} 
 To avoid a Landau pole in the renormalization group equation 
 we require $\tilde{u} \leq 1$. Since we also require $\tilde{u}>0$ so that $\alpha>0$, which is needed for (\ref{ZX}) and (\ref{ZYX}), 
 the value $\tilde{u}=1$ is uniquely selected. Then $\alpha_3$ runs only between $m_Z$ and $m_Y$ and its value at $m_X$ is approximately the same as at 
 $m_Z$. Then $m_w$ is at least a factor of $10^2$ larger than $m_s$, and
  it is therefore consistent to identify 
 the string scale  with the unification scale.  The values caculated in (\ref{ZX}) and (\ref{ZYX}) are (just about) consistent with the 
 observable supersymmetry breaking arising from gravitational interactions with an anti-brane sector.
 
 The situation with respect to quark and lepton mass hierarchies is the same as for the model discussed by Aldazabal {\it et al}\cite{AIQU}. In particular,
 superpotential terms coupling the chiral superfields $L,e^c_L$ and $H_d$ are forbidden at tree level, and to all orders string perturbation theory. 
 Such couplings are allowed, of course, by conservation of weak hypercharge, but {\it not} by conservation of $Q_2$, for example. As discussed in \S 5, 
 such $U(1)$ symmetries are expected to survive as {\it global} symmetries in Type I/IIB theories. However, the global and other symmetries do allow
  the coupling  of $L$ and $e^c_L$ to a composite effective Higgs field constructed from $\bar{e}^c_L, H_u$ and some D7-brane chiral fields. Since 
  baryon number $B=\frac{1}{3}Q_3$, it is perturbatively conserved if conservation of $Q_3$ survives as a global symmetry. Thus the proton is stable. 
  On the other hand, lepton-number is not conserved, so another solution of this problem is required if such models are to survive. In any case, the 
  survival of global $U(1)$ symmetries brings other problems. Generically, we expect electroweak spontaneous symmetry breaking to break {\it both} the 
  local $SU(2) \times U(1)$ gauge symmetry and any global $U(1)$ symmetry. (This certainly applies to $Q_2$, since $H_u$ and $H_d$ have $Q_2=\pm 1$.)
  Spontaneous breaking of a global symmetry gives a massless Goldstone boson at the Lagrangian level, the ``axion'',
    which acquires a non-zero mass because the global $U(1)$  is anomalous. 
   Unfortunately, electroweak spontaneous symmetry breaking generates a 
   ``visible'' axion with a mass predicted to be of order $10$keV, and there is by now overwhelming evidence that such an axion does 
   not exist\cite{PDG}.
   
 In conclusion, we note that the bottom-up approach allows the construction in an attractively direct way of models with gauge group
  and matter content close to that of the standard model, but with an intermediate  unification scale. There remain non-trivial problems with the
  lepton mass hierarchy, lepton number non-conservation and a visible axion.

\section*{Acknowledgments}
This paper is based on the talk given by one of us (D.B.) at the (first) Cairo International Conference on High Energy Physics.
 It is a pleasure to thank 
 the organisers for their invitation to speak, and to congratulate them all, particularly Shaaban Khalil, for making the meeting so enjoyable. 
 We are grateful to Mark Hindmarsh for pointing out the problem with  visible axions in our model.

\end{document}